\documentstyle[12pt,fleqn]{article}
\renewcommand{\baselinestretch}{1.2}

\newcommand{\bea}{\begin{eqnarray}}
\newcommand{\beq}{\begin{equation}}
\newcommand{\eea}{\end{eqnarray}}
\newcommand{\eeq}{\end{equation}}
\newcommand{\nnu}{\nonumber}
\newcommand{\di}{\mbox{d}}
\newcommand{\spav}[1]{\parbox{1mm}{\vspace*{#1}}}

\begin{document}

\begin{titlepage}
\begin{flushright}
CERN-TH.6751/92
\end{flushright}
\spav{.5cm}\\
\begin{center}
{\Large\bf $T$ Violation Induced by Supersymmetry}\\
{\Large\bf in $t\bar{t}$ and $W^+W^-$ Physics}
 \spav{2cm}\\
{\large Ekaterina Christova}
\spav{1cm}\\
{\em Institute of
Nuclear Research and
 Nuclear Energy}\\
{\em Boul. Tzarigradsko
Chaussee 72, Sofia 1784, Bulgaria.}
\spav{1cm}\\and
\spav{1cm}\\
 {\large Marco Fabbrichesi}
\spav{1cm}\\
{\em CERN, Theory Division}\\
{\em CH-1211 Geneva 23, Switzerland}\\
\spav{1.5cm}\\
{\sc Abstract}
\end{center}
$T$-odd correlations of polarizations and momenta
provide a promising testing ground for new physics beyond
the standard model.
We estimate the contribution of the minimal
supersymmetric extension of the standard model
to two such observables: in the production of
$t\bar{t}$, we look for a  term  proportional to
\mbox{$\mbox{\bf J}_t \cdot \left(
\mbox{\bf p}_q \times  \mbox{\bf p}_{t} \right)$}---where {\bf J}$_t$ is
the polarization of the $t$ quark and {\bf p}$_{q,t}$
are the momenta of the initial
and final particles---and find that it is of the order of
 $10^{-1} \times (\alpha_s/\pi)$. In the production of $W^+W^-$, we look for
a term proportional to \mbox{$\mbox{\bf E}_W \cdot \left(
\mbox{\bf p}_q \times  \mbox{\bf p}_{W} \right)
\left( \mbox{\bf p}_q \cdot
\mbox{\bf E}_W \right)$}---where {\bf E}$_W$ is the transverse
polarization of W--- to find that it can be as large
 as $10^{-1} \times (\alpha_w/\pi)$.

\vfill
\spav{.5cm}\\
CERN-TH.6751/92\\
Revised version, May 1993.

\end{titlepage}

\newpage
\setcounter{footnote}{0}
\setcounter{page}{1}

{\bf 1.} Observables which, in a given cross section, are
made out of an odd number of momenta and polarizations
change sign under time reversal.

If we assume that $CPT$ invariance holds, such a $T$-odd correlation can arise
either because of final state interactions~\cite{FSI}
 or because of a violation of $CP$
invariance.

The former is a consequence of the unitarity of the $S$ matrix
and carries no new dynamical information. It is a background
that can be subtracted by
taking the difference between the process we are interested in
 and its $CP$
conjugate~\cite{Golowich}. This way,
the truly (that is, $CP$-odd) time-reversal-violating observable is
isolated.

Such observables are negligible  in the standard
model---where the only possible $CP$-odd
source of such a $T$-odd correlation is in the
Kobayashi-Maskawa quark-mixing matrix, the effect
of which is, however, suppressed by
the unitarity of the matrix itself---and,
for this reason, they provide a promising testing ground
for physics beyond the standard model~\cite{others,Kane,us}.

New physics may be unveiled
 either because some of the final states are made of new
particles or because of its effects in the radiative corrections to the
amplitude of the process. We follow this latter path and include one-loop
corrections in the framework of the minimal supersymmetric extension of the
standard model~\cite{MSSM}. This model is  of interest here
inasmuch as time reversal
invariance can be
violated to a larger degree
than  in the standard model because of
the presence of coupling strengths that cannot be made real
 by a suitable redefinition
of the particle fields.

We consider two processes which should  give rise to
measurable $T$-odd and $CP$-violating observables.

The first one is the  production of $t\bar{t}$ pairs in hadron
$q\bar{q}$ collisions and in $e^+e^-$ annihilation:
\beq
e\bar{e} \rightarrow t\bar{t} \, ,  \label{1}
\eeq
in which we look for terms in the cross section proportional
to
\beq
\frac{\mbox{\bf J}_t \cdot \left( \mbox{\bf k} \times \mbox{\bf p} \right)}
{| \mbox{\bf k} \times \mbox{\bf p} |} \, ,
\label{a}
\eeq
where {\bf J}$_t$ is the
polarization vector of one of the produced $t$ quarks,
{\bf k} and {\bf p} are two vectors characterizing the scattering
plane---hereafter chosen to be {\bf k}, the
center-of-mass momentum of the colliding pair,  and {\bf p},
 the momentum of the final $t$ quark\footnote{Other $CP$-odd
observables have been recently computed within the minimal supersymmetric
extension of the standard model~\cite{Bern}.}. Because the
vector product
 $\mbox{\bf k} \times \mbox{\bf p}$ defines a vector perpendicular to the
 production plane,  only the component of the $t$ quark polarization
 that is transverse to this plane can appear in the
 correlation~(\ref{a}); therefore, {\bf J}$_t$ will  denote such
 transverse polarization only.

 As pointed out before, a transverse polarization of the $t$-quarks can
 be generated either in interactions between the final-state
 fermions, through the imaginary part of the loop integrals the
 amplitude---the so-called unitarity background---or by
$CP$-violating phases in
 the Lagrangian. After a $CP$-transformation, the transverse
 polarization of the $t$ quark and the $\bar{t}$ anti-quark
which originate
 in the final-state interactions should be equal while they should point
 in opposite directions in the case where
they arise from $CP$-violating pieces in
 the Lagrangian.

 Let us then suppose that we can measure the transverse polarizations of
 $t$ and $\bar{t}$ in future collider experiments, using, for instance,
 the method discussed in~\cite{Kane}. A comparison between
  the two transverse polarizations would make it possible to
remove the unitarity
 background because any
difference between them would imply
$CP$-violation in the $t\bar{t}$ production
process.

 It is also possible to single out the $T$-odd, $CP$-violating
 contribution by a direct estimate of the degree of transverse
 polarization due to final-state interactions. The QCD one-loop
 contribution, governing the leading behavior in the standard model, has
 been computed in~\cite{Kane}. An enhancement of the predicted
 polarization effect would then be a signal of new physics.

The chiral structure of
the supersymmetric amplitude is such that~(\ref{a})
 is proportional
to  the mass of the
$t$ quark.
It is for this reason that
the $t$ quark, with its large mass~\cite{CDF}, is such
a good candidate for observing a non-vanishing value of~(\ref{a}).
For $m_t$ in the present experimental range, and supersymmetric masses around
200 GeV,
the supersymmetrical correction is of the same
 order as a
one-loop
radiative correction within the standard model, which we can take
to be typically of the order of $10^{-1} \times (\alpha/\pi)$, where
$\alpha$ can be either $ g^2_s/4\pi$ for a strong coupling correction or
$g^2_w/4\pi$ for a weak correction.

The second process that  we consider is the production of $W^+W^-$ pairs in
\beq
q\bar{q}\rightarrow W^+W^-
 \quad \mbox{or} \quad  e\bar{e} \rightarrow W^+W^- \, ,
\eeq
in which we estimate  the term in the cross section proportional to
the correlation
\beq
\frac{\mbox{\bf E}_W \cdot \left(
\mbox{\bf k} \times  \mbox{\bf p} \right)}{|\mbox{\bf k} \times  \mbox{\bf p}|}
\frac{ \mbox{\bf k} \cdot
\mbox{\bf E}_W}{|\mbox{\bf k}|} \, ,   \label{b}
\eeq
where {\bf E}$_W$ is the transverse polarization of one of the final
vector bosons; as before, {\bf k} and {\bf p}  are, respectively,
 the center-of-mass momentum
of the colliding pair and of the W's. {\bf E}$_W$ has two components, one
parallel to the reaction plane which gives a non-vanishing
contribution to the
scalar product  with {\bf k}, and one transverse to such a plane and
appearing in the  triple product in~(\ref{b}). These two components of
the  transverse polarization of $W$ must  both be different from zero in order
for the observable~(\ref{b}) to be measurable.

As in the case of the $t$ quarks, the correlation~(\ref{b})
 takes opposite signs for the polarization of, respectively, $W^+$ and
 $W^-$ if it originates from $CP$-violating phases in the
 Lagrangian and the same sign if it comes from the unitarity background.

 The $T$-odd correlation~(\ref{b}) arising from the $CP$-violating part of
the supersymmetric Lagrangian turns out to be of  order
$10^{-2} \times \left( \alpha_w /\pi \right)$. In a narrow range of
scattering angles in the backward direction, it grows of one order of
magnitude to become of the
same order of a  one-loop radiative correction within the standard model.

Both the $t$ quark and the $W$ boson
 will become copiously available  as new accelerators
(the LEPII, LHC and SSC) come into operation. This will make possible
not only
a detailed study of their properties, but also an efficient test of
the possible non-vanishing of the
$T$-odd correlations (\ref{a}) and (\ref{b}).

\spav{1.5cm}\\
{\bf 2.} Let us first  fix our notation by writing those
parts of
 the minimal supersymmetric
extension of the standard model we need.

We neglect generation mixing. Hence, only three
terms in the supersymmetric Lagrangian  can give
rise to $CP$-violating phases  which cannot be rotated
away~\cite{phases}: The superpotential
contains a complex coefficient $\mu$ in the term bilinear
in the Higgs superfields. The soft supersymmetry breaking operators
introduce two further complex terms, the gaugino masses
$\widetilde{M}_i$ and the left- and right-handed squark mixing
term $A_q$. We
consider only the latter two, which are carried by truly supersymmetric
particles, and leave out the additional contribution of the Higgs sector.

 The
 squark  mass eigenstates $\tilde{q}^j_{\alpha,n}$ are related to the weak
eigenstates $\tilde{q}^j_{\alpha,L}$ and $\tilde{q}^j_{\alpha,R}$
through the mixing matrix:
\bea
\tilde{q}^j_{\alpha,L} & = & \exp (-i\phi_{A_q}/2) \left[ \cos \theta\:
\tilde{q}^j_{\alpha,1} + \sin \theta \:   \tilde{q}^j_{\alpha,2} \right] =
\sum_m a^L_m \tilde{q}^j_{\alpha,m} \\
\tilde{q}^j_{\alpha,R} & = & \exp (i\phi_{A_q}/2) \left[ \cos \theta\:
\tilde{q}^j_{\alpha,2} - \sin \theta \:   \tilde{q}^j_{\alpha,1} \right] =
\sum_n a^R_n \tilde{q}^j_{\alpha,n}
\eea
where
\beq
\tan 2\theta = \frac{2|A_q|m_q}{(L^2 - R^2)\widetilde{m}}
\eeq
and
\beq
A_q \widetilde{m}
m_q = \xi_q v_2 + \mu^* h_q v_1 \qquad A_q = |A_q| \exp
i\phi_{A_q} \, . \eeq
$\widetilde{m}L$ and $\widetilde{m}R$ are the squark mass parameters,
$v_i$ the vacuum expectation values of the Higgses, and $\xi_q$ the
coefficient in the cubic term of the soft breaking operator. The
diagonalization of the squark masses gives the eigenvalues
\beq
\widetilde{m}_{1,2}^2 = \frac{1}{2} \left\{ (L^2 + R^2)\widetilde{m}^2
 + 2m_q^2 \mp
\left[ (L^2 - R^2)^2 \widetilde{m}^4 + 4m_q^2|A_q|^2
\widetilde{m}^2 \right]^{1/2}
\right\} \, . \label{m12} \eeq

 The gluino majorana mass
\beq
\tilde{M}_g = \tilde{m}_g \exp (i\phi_g)
\eeq
gives an additional phase shift once it has been rotated into the
interaction to make the masses  real.

The neutralino  mass eigenstates $\tilde{\chi}^0_i$ are 4-component Majorana
spinors, whose left-handed components are related to the weak interacting
two-component spinor fields
 \beq
\psi_j^0 = \left( -i\lambda ',-i\lambda^3,\psi_{H_1}^0,\psi_{H_2}^0
 \right)
\eeq
through the $4\times 4$ neutralino mixing matrix $N$:
\beq
\tilde{\chi}^{0}_{iL} = N_{ij} \, \psi^0_{jL} \, , \qquad i=1,2,3,4\, .
\eeq
 The
chargino two-component mass eigenstates $\chi^{\pm}_{i}$ are defined by
\beq
\chi^{+}_{i} =  V_{ij} \psi^{+}_{j}  \qquad \mbox{and} \qquad
\chi^{-}_{i}  =  U_{ij} \psi^{-}_{j} \, , \qquad i=1,2 \, ,
\eeq
where
\beq
\psi^{+}_{j}   =  \left( -i\lambda^+ , \psi_{H_2}^{+} \right) \qquad
\mbox{and} \qquad
\psi^{-}_{j}  =  \left( -i\lambda^- , \psi_{H_1}^{-} \right)
\eeq
are the  weak interacting spinor fields.
 Here, $V$ and $U$ are
$2\times 2$ unitary matrices that diagonalize the wino-higgsino mass
matrix. The 4-component chargino field is
\beq
\tilde{\chi}^{+}_{k}=\left( \chi^{+}_ {k},\bar{\chi}^{-}_{k}\right),
\eeq
whose left and right-handed components are expressed in terms of the two
actually independent
matrices $U$ and $V$.
Both the neutralino $N$ and the chargino $V$ and $U$ mixing matrices are
determined by the supersymmetry breaking mechanism and
contain in general $CP$-violating phases~\cite{MSSM}.

We can construct the diagrams
of Figs. 1 and 2 by means of the supersymmetric Lagrangian~\cite{MSSM}.
In the computation we present in this letter,
we make use, in addition to
 those of the standard model, of  the following six terms:
\bea
L_{\tilde{t}\tilde{t}V^i} &=& i  g_i V^i_\mu \left(
\epsilon
^i_L \tilde{t}^*_L
\stackrel{\leftrightarrow}{\partial^\mu} \tilde{t}_L  +
\epsilon^i_R \tilde{t}^*_R
\stackrel{\leftrightarrow}{\partial^\mu} \tilde{t}_R  \right) \label{L1}
 \, ,  \\
L_{Z\tilde\chi^+\tilde\chi^-} &=&  g_Z
Z_\alpha \sum_{i,j} \bar{\tilde\chi}^+_i \gamma^\alpha \left [Q^L_{ij}
(1-\gamma_5 )
+ Q^R_{ij} (1 + \gamma_5) \right ] \tilde\chi^+_j  + h.c.\, , \label{L2}\\
L_{\gamma\tilde\chi^+\tilde\chi^-} &=&-eA_{\alpha}\bar{\tilde\chi}
^+_i\gamma^\alpha\tilde\chi^+_j \: \delta_{ij} \\\
L_{W^-\tilde\chi^+\tilde\chi^0} & = & \frac{g}{2}
W_\alpha \sum_{k,i} \bar{\tilde\chi}^0_k \gamma^\alpha \left[ O^L_{ki}
(1-\gamma_5 )
+ O^R_{ki} (1 + \gamma_5) \right] \tilde\chi^+_i  + h.c. \, , \\
L_{\tilde{q}q \tilde{g}} &=& \frac{g_s}{\sqrt{2}} T^a_{jk}
 \sum_{\alpha=t,b}
\left[  \bar{\tilde{g}}_a (1-\gamma_5) q_\alpha ^k
\Gamma_{L*}^m \tilde{q}^{j*}_{\alpha ,m} + \bar{q}_\alpha ^j (1+\gamma_5)
\tilde{g}_a \Gamma_{L}^m\tilde{q}^{k}_{\alpha ,m}  \right. \nnu \\ & & - \left.
\bar{\tilde{g}}_a (1+\gamma_5) q_\alpha ^k \Gamma_{R*}^n
\tilde{q}^{j*}_{\alpha
,n} - \bar{q}_\alpha ^j (1-\gamma_5) \tilde{g}_a
\Gamma_{R}^n\tilde{q}^{k}_{\alpha ,n} \right]
 + h.c.\label{L3}  \\
L_{Z\chi^0\chi^0} & = & \frac{g_Z}{2}
Z_\alpha \sum_{j,i} \bar{\tilde\chi}^0_i \gamma^\alpha \left[ R^L_{ij}
(1-\gamma_5 ) + R^R_{ij} (1+\gamma_5) \right] \chi^0_j + h.c.
\, .
\eea

A few comments on the notation:
The index $i$ in (\ref{L1}) keeps track of whether
the scalar-quark vertex is coupled to $\gamma$ or $Z$. We have
\bea
g_\gamma & = &  g\sin\theta_W =
e \, ,\qquad \qquad g_Z=g/2\cos \theta_W \, ,\nnu \\
\epsilon^\gamma_L & = & \epsilon^\gamma_R =2/3 \, , \nnu \\
\epsilon^Z_{L} & = & 1- (4/3) \sin^2\theta_W \, ,
\qquad \qquad \epsilon^Z_{R}  = - (4/3) \sin^2\theta_W  \, .
\eea
Moreover, in (\ref{L2})-(\ref{L3}) we have used the notation
\bea
\Gamma_{L}^m  & = & a_m^{L*}\exp ( -i\phi_g) \nnu \\
 \Gamma_{R}^n  & = &a_n^{R}\exp (-i\phi_g) \nnu \\
O^L_{ki} & \equiv & -\frac{1}{\sqrt{2}} N_{k4}V^*_{i2} +
N_{k1}V^*_{i1} \, , \qquad Q^L_{ij} \equiv
-V_{i1}V^*_{j1} - \frac{1}{2} V_{i2}V^*_{j2} + \sin^2\theta_W  \, ,\nnu \\
O^R_{ki} & \equiv & -\frac{1}{\sqrt{2}} N_{k3}^*U_{i2} +
N_{k1}^*U_{i1} \, ,  \qquad Q^R_{ij} \equiv
-U_{i1}U^*_{j1} - \frac{1}{2} U_{i2}U^*_{j2} + \sin^2\theta_W  \nnu  \\
R^L_{ij} & \equiv & \frac{1}{2} N_{i3}N_{j3}^* + \frac{1}{2}
N_{i4}N_{j4}^*   \, ,\qquad R^R_{ij} =  R^{L*}_{ij} \, .
 \eea
In~(\ref{L1}) and below, $\tilde{g}_a$ are the gluinos
and
$\tilde{t}_{L,R}$ are the scalar left- and right-handed  partners
of the $t$ quark field.

\spav{1.5cm}\\
 {\bf 3.} We  turn now to estimating the observable~(\ref{a})
 in the
production of $t\bar{t}$ in the minimal supersymmetrical extension of the
standard model. It can receive a contribution from several diagrams
(Fig.1). However, as long as we are not committed to a particular model
of supersymmetry breaking, it is sufficient to consider just a class of
them. We take only the gluino Penguin diagram, depicted in Fig. 1(a)
which is slightly enhanced by a factor $\alpha_s/\alpha_w$ with respect
to the neutralino and chargino diagrams.

The gluino Penguin diagram gives a contribution to the following
$CP$-violating and helicity flipping term:
\beq
\bar{u} (p) \Gamma^i_\alpha u(-p') =
i g_i \bar{u} (p)\:  P_\alpha\: \gamma_5 u(-p') \, {\cal B}^i
\qquad \qquad i=\gamma, Z \, ,
\eeq
where $p$ and $p'$ are the momenta of, respectively,
 $t$ and $\bar{t}$ and $P=p-p'$. The
${\cal B}^i$ are  real functions given by:
\bea
{\cal B}^i  &= &-2g^2_s  \widetilde{m}_g \left(
\epsilon^L_i +
  \epsilon^R_i\right) \nnu \\
& & \times   \sum_{n,m}  \mbox{Im} \Biggl\{
\left(
\Gamma^{R*}_n\Gamma^{L}_n -    \Gamma^{L*}_m\Gamma^{R}_m \right) \left(
\Gamma^{R*}_m\Gamma^{R}_n + \Gamma^{L*}_m\Gamma^{L}_n \right)
\Biggr\}  \Biggl[
2a^{m,n} + I^{m,n} \Biggr] \, . \label{BB}
\eea
In eq.~(\ref{BB}), $I^{m,n}$ and $a^{m,n}$ are defined
by means of the loop integrals as follows:
\beq
I^{m,n} = \int \frac{\di ^4 k}{(2\pi)^4} \frac{1}{k^2 -
(\widetilde{m}_g)^2} \frac{1}{(k-p')^2 - (\widetilde{m}_{n})^2}
\frac{1}{(k+p)^2 - (\widetilde{m}_{m})^2} \, ,  \label{I1}
\eeq
and
\bea
I_{\alpha}^{m,n} &=& \int \frac{\di ^4 k}{(2\pi)^4} \: k_\alpha
\: \frac{1}{k^2 -
(\widetilde{m}_g)^2}
\frac{1}{(k-p')^2 - (\widetilde{m}_{n})^2}
\frac{1}{(k+p)^2 - (\widetilde{m}_{m})^2} \nnu \\
&= & a^{m,n} P_\alpha + b^{m,n} q_\alpha \, ,     \label{I2}
\eea
where $q=p+p'$.

The matrix element for the process contains, beside the tree-level
standard model
term,  the following $CP$-violating amplitude:
\bea
&& i{\cal N}\left[ \frac {e^2}{s}\,\bar{u} (-k') \gamma_\alpha  u(k)\,\,
\bar{u}(p)\:iP^\alpha {\cal B}^\gamma \gamma_5 u(-p') \right.
 \label{22} \cr
& & +\:\left. \frac{g_Z^2}{s-M_Z^2} \bar{u}(-k') \gamma_\alpha
\left( c_V + c_A \gamma_5 \right) u(-k) \,
\,\bar{u}(p)\: iP^\alpha {\cal B}^Z \gamma_5 u(-p')\right ] \nnu \\
& & \times \:
(2\pi)^4\delta (p+p'-k-k') \label{amp} \, .
\eea

In eq.~(\ref{22}), $c_V \equiv -(1/2) +2\sin^2\theta_W$ and $c_A \equiv 1/2$,
 $\cal N$ is the usual factor containing the normalization of the states.

The cross section can now be computed. It contains the standard model
tree-level part $\di \sigma_0^{t\bar{t}}
 / \di \Omega$~\cite{Renard} together with the
observable~(\ref{a}) which arises in the interference with the one-loop
supersymmetric amplitude~(\ref{amp}). It can be written as
\beq
\frac{\di \sigma}{\di \Omega} =
\frac{\di \sigma_0^{t\bar{t}}}{\di \Omega} \left( 1 + D_t\,
\frac {\left ( \mbox{\bf J} \cdot{\mbox{\bf p}}\times \mbox{\bf k}\right
)}
{\vert\mbox{\bf p}\times\mbox{\bf k}\vert }
\right) \, .
\eeq
Here $\mbox{\bf J}$ is the unit polarization vector perpendicular to the
production plane, defined by ${\bf J}_t=D_t\cdot\mbox{ \bf J}$.
The degree of transverse polarization  $D_t$,
 which gives the magnitude of the $T$-odd,
$CP$-violating effect we are after is in evidence.
 Our computation yields:
\bea
D_t & =& - \left(\frac{1}{{\cal T}_0}\right)
\sqrt{s}\beta\sin\vartheta\tan^2\theta_W \frac{s}{s-M_Z^2}\:
 \\
&  & \times \: \left\{ c_A \left[ g_V \,{\cal B}^\gamma +
\left( \frac{2}{3}
-\frac{c_Vg_V}{2}
 \frac{1}{\sin^2\theta_W\cos^2\theta_W} \frac{s}{s-M_Z^2} \right)
 \,{\cal B}^Z \right] \right. \nnu \\
& & \left. -\: \beta\cos\vartheta g_A \left[ c_V \,{\cal B}^\gamma -
\left(
\frac{c_V^2 + c_A^2}{4}
 \frac{1}{\sin^2\theta_W\cos^2\theta_W} \frac{s}{s-M_Z^2} \right)
 \,{\cal B}^Z \right] \right\} \, , \nnu \label{Dt}
\eea
where
\beq
{\cal T}_0 = 16s \frac{\di \sigma_0^{t\bar{t}}/\di \Omega}
{\alpha^2_W\beta}
\, .
\eeq
In (\ref{Dt}) $g_V = 1/2 - (4/3)\sin^2 \theta_W$,
$g_A =1/2$ and $\beta = \vert\mbox{\bf
p}\vert/E$. The scattering angle is denoted by $\vartheta$.

The masses of the  squarks can be taken as equal
once we have factorized out the overall dependence on
them  of the integrals,
so that the sum in (\ref{BB}) does not vanish by the orthogonality
of the mixing matrices $\Gamma^{L,R}$. This factorization gives
 a factor
\beq
\frac{\widetilde{m}_2^2 - \widetilde{m}_1^2}
{\widetilde{m}_1^2\widetilde{m}_2^2} \simeq \frac{2|A_t|m_t}
{\widetilde{m}_1^2\widetilde{m}_2^2} \widetilde{m} \, ,
\eeq
where we have assumed for simplicity in (\ref{m12}) $L=R$. In the same
approximation $\sin \theta \cos \theta =1/2$.
All supersymmetrical masses can now be assumed to be
of the same order, and their value denoted  by $\widetilde{M}$. We are
thus left with only two arbitrary supersymmetric parameters: one
$CP$-violating phase and one supersymmetric mass.  Moreover, even though
expression (\ref{Dt}) is correct both below and above the threshold for
production of super particles, we are more interested in the case in
which no superparticles are produced, i.e.
to the regime
\beq
2m_t  < \sqrt{s} < 2\widetilde{M} \, . \label{regime}
\eeq

 Under these assumptions, the integrals~(\ref{I1}) and (\ref{I2})
can be computed  to yield
\beq
I^{m,n} = I \equiv \frac{1}{4\pi^2s} \widetilde{I}
\eeq
where
\beq
\widetilde{I} \equiv \int_{0}^{1} \di x \frac{1}{\sqrt{\Delta}}
\arctan \frac{(1-v)\,(1-x)}{\sqrt{\Delta}} \, ,  \label{II1}
\eeq
and
\beq
\Delta \equiv (1-v) \left[ (1-x)^2 -v(1-x) +u \right] \, .
\eeq
The two parameters
\beq
v \equiv 4m_t^2/s \quad \mbox{and} \quad u \equiv 4\widetilde{M}^2/s \, .
\eeq
are such that $v<1$ and $u>1$ because of (\ref{regime}).

Similarly,
\beq
a^{m,n} = a \equiv \frac{1}{4\pi^2s} \left( -\frac{1}{2} \widetilde{a}
\right) \, ,
\eeq
where
\beq
\widetilde{a} \equiv \int_{0}^{1} \di x \frac{1-x}{\sqrt{\Delta}}
\arctan \frac{(1-v)\,(1-x)}{\sqrt{\Delta}} \, , \label{II2}
\eeq

Hence, we obtain that
\beq
{\cal B}^i
 \, \simeq -8 (\epsilon_L^i+\epsilon_R^i) \frac{\alpha_s}{\pi}\sin
\delta_{CP} \frac{m_t}{s}\,
  (\widetilde{I}-\widetilde{a})  \label{B} \, ,
\eeq
where we have denoted by $\sin \delta_{CP}$ the phase $|A_t|\sin
(\phi_{A_t} - \phi_g )$.

 The value of $\tilde{{\cal B}}^i$ determines the
order of magnitude of the effect. By assuming maximum
$CP$-violation ($\sin \delta_{CP}=1$), the effect can be made of the order of
magnitude of a  one-loop radiative correction.

We have  evaluated the integrals (\ref{II1}) and (\ref{II2})
numerically to obtain the dependence of
the $T$-violating
coefficient $D$
in
\beq
D_t = \left( \frac{\alpha_s}{\pi} \right) \: D \:
\sin \delta_{CP}
\eeq
 on the scattering angle $\cos
\vartheta$,
 for different choices of
 $\sqrt s$, $m_t$ and $\widetilde{M}$ (see Fig. 3).

Figs. 3(a)-(b) show that the size of $D$
is proportional to $m_t$, becoming smaller
for  a larger supersymmetric mass scale $\widetilde{M}$.

The energy dependence  has a kinematic origin. $D_t$
is normalized by the
tree-level cross section ${\cal T}_0$. The enhancement in the
backward direction (see Fig. 3) is due to the fact that
 ${\cal T}_0$ is
smaller there. In the same backward direction,
${\cal T}_0$ decreases as the center-of-mass energy $\sqrt{s}$ is
increased, and therefore $D$ is made bigger (see Fig. 3(c)).

For $\sqrt{s} = $ 280 GeV,  $\widetilde{M} = 150$ GeV and $m_t =$ 130
GeV we have $D \simeq .1$, the value that gives the
 estimate quoted in
the abstract.

\spav{1.5cm}\\
 {\bf 4.} Our second example,
the observable (\ref{b}) in the production of $W^+W^-$,  can be
estimated as follows.

The most general Lorentz invariant coupling of a neutral current
$\Gamma^\mu$ to a pair
 of conjugate vector bosons $W^+_\alpha\,W^-_\beta$
 can be parameterized in terms of ten form
factors $f_i$ (see, for instance, \cite{GG}, the notation of which we
follow). Four of them are $CP$-violating and potentially give rise to the
correlation (\ref {b}). However, as we want to obtain just an estimate of
the effect,
  we only need
consider one, for example, $f_6$ that is defined by the $WWV^i$ vertices
as follows:
\beq
\Gamma_\mu^i = \tilde {g}_i\,
f_6^i \epsilon_{\mu\sigma\alpha\beta}\:
q^\sigma E^\alpha _- E^\beta _+ \, \qquad \qquad i=\gamma, \,Z \, ,
\label{f5}
\eeq
where
$\tilde{g}_\gamma =e$ and $\tilde{g}_Z = e\cot\theta_W$.
In (\ref{f5}) the $E_\pm$ are the $W^\pm$-polarization 4-vectors,
$q = p+ p'$; $p$ and $p'$ being the momenta of, respectively, $W^-$
and $W^+$.

The vertex (\ref{f5}) corresponds to an effective gauge non-invariant
operator
 \beq
i\kappa^i W_\mu^{\dag} W_\nu \tilde{V}^{\mu\nu}_i +
\frac{i\lambda^i}{M_W^2} W^{\dag}_{\lambda\mu}W^\mu_\nu
\tilde{V}^{\nu\lambda}_i \, , \eeq
where $\tilde{V}_i^{\mu\nu}$ stands for either the photon or the $Z$ dual
field strength and $f^i_6 = \kappa^i - \lambda^i$. This operator
originates by spontaneous symmetry breaking from an higher dimensional
gauge invariant operator~\cite{Gavela}. It  gives rise to an electric
dipole moment for the $W$'s~\cite{Keung}.

There are four types of loop diagrams which can give a contribution
to~(\ref{f5})---see Fig. 2. However, as we did for the $t$ quark case,
 it
 is sufficient to consider just a class of them. We take only the vertex
diagrams,
depicted in Figs. 2(a) and 2(b), neglecting the box-diagrams.

Since both the supersymmetric masses and the imaginary phases have no definite
value, we can also make the further simplification of
considering the case $i=j$. This leaves us with only one diagram, the one in
Fig. 2(a), the other one, in Fig. 2(b), being real under these assumptions.
The $CP$-violating phases of the remaining diagram  arise only
from the $\tilde{\chi}^0_k \tilde{\chi}^+_i W^-$-vertex, the diagonal
couplings
 $Q_{ii}^L$ and  $Q_{ii}^R$ in the $\tilde{\chi}_i\tilde{\chi}_i V$ vertex
being real.

A straightforward algebraic manipulation gives
\bea
f_6^Z & = & -\frac{2g^2}{\cos^2\theta_W} \sum_{k,i} \widetilde{m}^0_k
\widetilde{m}_i \mbox{Im} \, (O^L_{ki}O^{R*}_{ki}) (Q^L_{ii}+Q^R_{ii})
I_{k} \nnu \\
f_6^\gamma &  = & 4\,g^2 \sum_{k,i} \widetilde{m}^0_k
\widetilde{m}_i \mbox{Im} \, (O^L_{ki}O^{R*}_{ki})
I_{k} \, ,
\eea
where $I_{k}$ is the same integral as the one defined in (\ref{I1}) but
for the
masses of the  squarks being replaced by the masses $\tilde {m}_i$
 of the charginos and the mass of the gluino by the masses
$\widetilde{m}_0^k$ of the neutralino.

Correlation (\ref{b}) in the relevant cross section arises from
the interference of the one-loop amplitude
\bea
& & i{\cal N} \left[ \frac{g_\gamma}{s}  \:\Gamma^\gamma_\mu\:
\bar{u}(-k') \gamma^\mu  u(k)  \label{122}\right. \\
 & & + \left. \frac{g_Z}{s-M_Z^2}  \:\Gamma^Z_\mu \:
\bar{u}(-k') \gamma^\mu \left( c_V + c_A \gamma_5 \right)
 u(k)\right] \,(2\pi )^4\delta (p+p'-k-k') \nnu \, ,
\eea
where $\Gamma^i$ are given by (\ref{f5}),
with the  tree-level amplitude corresponding to diagrams with
  $Z$,
photon  and  neutrino being exchanged.

We assume  the polarization of $W^-$ to be fixed and we take
it to be transverse to the momentum of $W^-$. In this case $E_W$ can be chosen
to have space components only, $E_W=(0,{\bf E}_W)$. ${\bf E}_W$ can be
decomposed into two real polarization vectors: one perpendicular to the
momentum of $W^-$ and parallel to the reaction plane and the other
transverse to the reaction plane. In the center-of-mass system we have also
$(E_W\cdot p)=0$ and $(E_W\cdot q)=0$. This way, after summing
over the polarizations of the other vector boson
$W^+$, we obtain the dependence of the cross
section on the transverse  polarization of $W^-$. It
 can be written in the form:
\beq
\frac{\di \sigma}{\di \Omega} =
\frac{\di \sigma_0^{W^+W^-}}{\di \Omega}
\left ( 1 + D_W \:\frac{\left (\mbox{\bf E}_{W^-} \cdot
\mbox{\bf k}
\times \mbox{\bf p}\right )}{\vert \mbox{\bf k}\times \mbox{\bf p}\vert } \:
\frac{\mbox{\bf E}_{W^-}
\cdot \mbox{\bf k}}{\vert \mbox{\bf k}\vert } \right) \, ,
\eeq
where  the tree-level standard model contribution
$\di \sigma_0^{W^+W^-}/\di
\Omega$~\cite{Alles} has been factorized out to put
 in evidence the $T$-odd, $CP$-violating coefficient $D_W$, which
determines the magnitude of the effect.

This coefficient is now
\bea
D_W & = & \frac{\beta\sin\vartheta}{{\cal C}_0}
\left\{ f_6^\gamma \sin^2 \theta_W \left[ \frac{4s}{M_W^2} \left(
\sin^2 \theta_W + \frac{c_V}{2} \frac{s}{s-M_Z^2} \right) -
\frac{s}{2t} \left(1 - \frac{2t}{M_W^2} \right) \right ] \right. \nnu \\
& & + \:  \frac{1}{2} f_6^Z \frac{s}{s-M_Z^2} \left[
\frac{4s}{M_W^2} \left(
c_V \, \sin^2 \theta_W + \frac{c_V^2 + c_A^2}{2} \frac{s}{s-M_Z^2} \right)
\right. \nnu
\\ & & \left. \left. -
\frac{s}{2t}(c_V - c_A) \left(1 - \frac{2t}{M_W^2} \right) \right ]\right\} \,
,\label{D}
\eea
where
\beq
{\cal C}_0 = 16\, s \,
\frac{ \di \sigma_0^{W^+W^-} /\di \Omega }{\alpha_W^2 \beta} \, ,
\eeq
and
\beq
t = M_W^2 - \frac{s}{2} \left(1 - \beta \cos \vartheta \right)  \, .
\eeq

Again we have a complicated expression that we want to simplify.
As we have done in the previous case, we take
the supersymmetric masses at about the same values $\widetilde{M}$ and
 assume a unique $CP$-violating phase:
\beq
\mbox{Im} \, (O^L_{ki}O^{R*}_{ki}) = \sin\delta_{CP}/2 \, ,
\eeq
as well as
\beq
Q^L_{ii} = Q^R_{ii} \simeq 1/2 \, .
\eeq

 The sum over the neutralino states is
non-zero only if their masses are different. This introduces a factor
\beq
\frac{(\widetilde{m}^0_j)^2 -(\widetilde{m}^0_k)^2}
{(\widetilde{m}^0_j)^2(\widetilde{m}^0_k)^2}
\simeq
\frac{M_Z \widetilde{M}}{(\widetilde{m}^0_j)^2(\widetilde{m}^0_k)^2}
 \label{supp}
\eeq
 in the simplest case of
mixing of only two neutralinos~\cite{MSSM}. Notice that this factor
suppresses the $CP$-odd vertices  as we take the
supersymmetric masses to infinity for a fixed neutralino mass difference
and gives zero if instead we take all masses to be
degenerate.

Under these assumptions we obtain the following expressions for
$f^\gamma_6$ and $f^Z_6$:
\bea
f^\gamma_6 &=& \frac{\alpha_W}{\pi}\sin\delta_{CP}\,
\frac{2 M_Z \widetilde{M}}{s}\widetilde{I}
\label {f1} \\
f^Z_6 &=&
-\frac{\alpha_W}{\pi}\sin\delta_{CP}\,\frac{M_Z\widetilde{M}}
{s\cos^2\theta_W}
\widetilde {I}, \label{f2}
\eea
where the integral $\widetilde {I}$ has already been defined in (\ref{II1}),
except that now
\beq
v \equiv 4M_W^2/s \, .
\eeq
We consider the energy range below the threshold for the production
of the superparticles.

 We can now use eqs.(\ref{f1}), (\ref{f2})
and (\ref{D}) to obtain an estimate of the effect. A numerical
evaluation of $D_W$ for the same choice of energies and supersymmetrical
mass as before is given in Figs. 4(a)-(b),
where we plot the coefficient $D$ defined by
\beq
D_W = \left( \frac{\alpha_w}{\pi} \right) \: D \: \sin \delta_{CP} \, .
\eeq

 For maximal  $CP$ violation,
and at $\sqrt{s} = 280$ GeV, the size of $D$ in most of the backward
direction  is about $10^{-2}$.
 It, however, increases quite steeply in the narrow
region where $0.90 \le \cos \vartheta  \le 0.95$ to become one order
of magnitude bigger, that is, the value quoted in the abstract. This
is also true for the tree-level cross section ${\cal C}_0$ which has a
peak in the forward direction because of the neutrino-exchange diagram.
Since this peak in ${\cal C}_0$ is made bigger by increasing the
center-of-mass energy, the size of $D_W$ is accordingly made smaller. As
the supersymmetrical scale $\widetilde{M}$ grows, the effect becomes
smaller.

Such a maximal $CP$ breaking is consistent with bounds coming from the
induced electric dipole moment of the neutron~\cite{Marciano}.

 \spav{1.5cm}\\
 {\bf 5.} The two $T$-odd and $CP$-violating observables we have computed
within
the minimal supersymmetric extension of the standard model turn out to be
rather large. They are of the same order of magnitude as, or
only one order of magnitude smaller than,
a one-loop radiative correction within the
standard model itself.
 They may provide independent bounds on the
$CP$-violating supersymmetric parameters (for
present limits see, for example, \cite{limits}) and clues
on new physics  if they are
measured and found to be different from zero.

\spav{1.5cm}\\
E.C. would like to thank J. Ellis and the Theory
Group at CERN for  their kind
hospitality. Her work has been partially supported by the Bulgarian National
Science Foundation, Grant Ph-16. M.F. thanks Lisa Randall and Fabio
Zwirner for helpful discussions.

\newpage
\renewcommand{\baselinestretch}{1}

\newpage

\spav{4cm}

{\bf Fig. 1:} The supersymmetric
 loop diagrams that give a contribution to the $T$-odd, $CP$-violating
observable in the production of $t\bar{t}$.

{\bf Fig. 2:} The supersymmetric
loop diagrams that give a contribution to the $T$-odd, $CP$-violating
observable in the production of $W^+W^-$.

{\bf Fig. 3:} The $T$-violating coefficient $D_t$ in the production
of $t\bar{t}$.

{\bf Fig. 4:} The $T$-violating coefficient $D_W$ in the production
of $W^+W^-$.

\end{document}